\newcommand{\be}{\begin{equation}}
\newcommand{\ee}{\end{equation}}
\newcommand{\Kepler}{{Kepler} }

\documentclass{pnastwo}

\url{http://TBD}
\copyrightyear{2013}
\issuedate{Issue Date}
\volume{Volume}
\issuenumber{Issue Number}

\begin{document}

\title{Architectures of Planetary Systems and Implications for their Formation}	

\author{Eric B. Ford\affil{1}{Center for Exoplanets \& Habitable Worlds, The Pennsylvania State University, 525 Davey Laboratory, State College, PA 16803, USA}\affil{2}{Department of Astronomy \& Astrophysics, The Pennsylvania State University, 525 Davey Laboratory, State College, PA 16803, USA}}

\contributor{Submitted to Proceedings of the National Academy of Sciences of the United States of America}

\maketitle

\begin{article}
\begin{abstract}
Doppler planet searches revealed that many giant planets orbit close to their host star or in highly eccentric orbits.  
These and subsequent observations inspired new theories of planet formation that invoke gravitation interactions in multiple planet systems to explain the excitation of orbital eccentricities and even short-period giant planets.
Recently, NASA's Kepler mission has identified over 300 systems with multiple
transiting planet candidates, including many potentially rocky
planets.  Most of these systems include multiple planets with sizes
between Earth and Neptune and closely-spaced orbits.  These systems
represent yet another new and unexpected class of planetary systems
and provide an opportunity to test the theories developed to explain
the properties of giant exoplanets.  Presently, we have limited knowledge about such planetary systems, mostly about their sizes and orbital periods.  
With the advent of long-term, nearly continuous monitoring by Kepler, the method of transit
timing variations (TTVs) has blossomed as a new technique for characterizing the gravitational
effects of mutual planetary perturbations for hundreds of planets.  TTVs can 
provide precise (but complex) constraints on planetary masses,
densities and orbits, even for planetary systems with faint host
stars.  In the coming years, astronomers will translate TTV observations
into increasingly powerful constraints on the formation and orbital evolution of
planetary systems with low-mass planets.  
Between TTVs, improved Doppler surveys, high-contrast imaging campaigns and microlensing surveys, astronomers can look forward to a much better understanding of planet formation in the coming decade.   
\end{abstract}


\abbreviations{AMD, angular momentum deficit; AU, astronomical unit; MMR, mean motion resonance; RV, radial velocity; STIPS, short-period tightly-packed inner planetary systems; TTVs, transit timing variations}


\dropcap{P}rior to the discovery of exoplanets, astronomers fine tuned theories of planet formation to explain detailed properties of the solar system.  The discovery of exoplanets has significantly increased our appreciation for the diversity of planetary systems in nature.  In this article we review the current understanding of the late-stages of evolution of planetary systems, focusing on observational constraints from radial velocity and transit observations.

\section{Hot Jupiters}
Radial velocity (RV) surveys have discovered over 400 planets, most with masses larger than that of Jupiter (http://www.exoplanets.org;  \cite{Butler06,Wright11}.  Many of the early RV discoveries were ``hot-Jupiters'', planets with orbital periods of up to several days and masses comparable to that of Jupiter or Saturn (e.g., \cite{MayorQueloz95}).  As the timespan of observations has increased, the median orbital period of RV-discovered planets has steadily increased to more than a year.  Now, we know that hot-Jupiers are a relatively rare outcome of planet formation.  Nevertheless, their existence and their orbital properties provide important clues to the planet formation process.  

\subsection{Disk Migration}
Prior to the discovery of hot-Jupiters, planet formation theories had been focused on explaining properties of the solar system
\cite{Lissauer93}.  The large masses of hot-Jupiters imply a substantial gaseous component and therefore rapid formation, before the protoplanetary disk is dispersed.  {\em In situ} formation of the rocky cores of hot-Jupiters is problematic due to the high temperature and low surface density of the disk so close to their host star.  Therefore, theorists explain hot-Jupiters starting with the formation of a rocky core at larger separations from the host star, followed by accretion of a gaseous envelope and migration to their current location.  The mechanism for migration is less clear.  There are two broad classes of models: a gradual migration through a disk \cite{KleyNelson12,Capobianco11} or the excitation of a large eccentricity followed by tidal circularization \cite{RasioFord96,WeidenschillingMarzari96,FabryckyTremaine07}.  In principle, planets could migrate through either a gaseous protoplanetary disk or a planetesimal disk.  If giant planet cores form beyond the ice line, then planetesimal disks would rarely be massive enough to drive the large-scale migration needed to form a hot-Jupiter.  On the other hand, a gaseous protoplanetary disk could power a rapid migration, perhaps too rapid \cite{KleyNelson08}.  It is unclear how the planets would avoid migrating all the way into the host star or why the migration would be halted to leave planets with orbital periods of $\sim$2-5 days \cite{FordRasio06}.  The apparent pile-up of hot-Jupiters with orbital periods of a few days could be the result of censoring, i.e., those planet that continued to migrate closer to their host star were either accreted onto the star, destroyed or reduced in mass due to stellar irradiation and mass loss.  Even in this case, a stopping mechanism must be invoked to produce the observed giant planets with orbital periods beyond $\sim$7 days, since tides rapidly become inefficient with increasing orbital separations.  

\subsection{Eccentricity Excitation plus Tidal Circularization}

Unlike disk migration, eccentricity excitation followed by tidal circularzation naturally explains the ``pile-up'' of hot-Jupiters at orbital periods of 2-7 days due to the rapid onset of tidal effects.  The large eccentricities required to initiate circularization could be generated in a variety of ways.  
The simplest scenario is planet-planet scattering, as it requires only one additional massive planet \cite{RasioFord96}.  In the case of two planets and no additional perturbers, the initial ratio of semi-major axes must be small enough to permit close encounters.  Such scenarios may arise naturally for giant planets due to rapid mass growth.  Alternatively, a system with more than two planets \cite{WeidenschillingMarzari96,LinIda97,AdamsLaughlin03,Chatterjee08,JuricTremaine08} will naturally approach instability on a much longer timescale.  Even for a simple three planet system, the timescale until close encounters can easily exceed ten million years \cite{Chatterjee08}, by which time the protoplanetary disk will have dissipated.  

Alternatively, a system of two (or more planets) may become unstable due to an external perturbation, such as secular interactions with a binary companion \cite{Holman97} or a stellar flyby \cite{LaughlinAdams98}.  This can lead to strong planet scattering, often following a prolonged phase of weaker interactions \cite{Malmberg11,Boley12,Kaib13}.  
Each close encounter between giant planets leads to a small perturbation to their orbits \cite{Katz97}.  Thus, it is typically a series of close encounters that excites the two planets' orbital eccentricities until one planet's pericenter is small enough to initiate tidal circularizaiton.  In practice, planetary systems that form two giant planets may well form additional massive planets, leading to a series of planet-planet scattering events and substantially increasing the probability for one to achieve a pericenter of just a few stellar radii.  

Another possible mechanism for eccentricity excitation involves secular (i.e., long-term) perturbations by one or more distant bodies (e.g., more planets, a brown dwarf or binary stellar companion).  If there is a large mutual inclination between the inner planet and the outer companion, then large eccentricities are possible, even for systems with large orbital period ratios \cite{MazehShahm79,Holman97,Ford00b,Naoz11}.  This could be particularly relevant for planets orbiting one member of a binary (or higher multiple) star system, even though the orbital period of the two stars is often much larger than the orbital period of the planet.  

For small mutual inclinations, exciting an eccentricity large enough to trigger tidal circularization requires a substantial angular momentum deficit (AMD) and a series of planets that serve to couple the inner giant planet to outer planets which are more likely to have a significant initial AMD \cite{ZakamskaTremaine04}.  
If the planets are widely spaced, then it is possible to construct initial conditions that lead to the inner planet's pericenter dropping to only a few stellar radii \cite{WuLithwick11}.  However, for more typical initial conditions, the secular interactions lead to close encounters between planets that result in collisions and/or ejections via planet-planet scattering \cite{LinIda97,Levison98,Chatterjee08,Nagasawa08,Matsumura10,MoeckelArmitage12}.  

\subsection{Distinguishing between Hot-Jupiter Formation Models}

In practice, nature may provide multiple migration mechanisms for forming hot-Jupiters.  For many years, observations provided little data to help distinguish between even the two broad classes of migration models.  The major breakthrough was the measurement of the Rossiter-McLaughlin effect for many transiting hot-Jupiters.  When the planet passes in front of a rotating star, the apparent radial velocity is perturbed due to the planet blocking a portion of the star that is rotating towards or away from the observer.  The Rossiter-McLaughlin signature measures the angle between the star's rotational angular momentum and the planet's orbital angular momentum (after projecting both onto the sky plane).  While many systems are well-aligned, a significant fraction of hot-Jupiters are severely misaligned \cite{Albrecht12}.  Misaligned systems, including nearly polar and even retrograde configurations \cite{Sanchis-Ojeda13}, arise naturally in scenarios that include eccentricity excitation plus tidal circularization \cite{Nagasawa08,Naoz11}.  While planet scattering and secular perturbations will only produce hot-Jupiters for a few percent of planetary systems \cite{FordRasio08,Naoz12}, this is consistent with the rate of hot-Jupiters observed, or at least a substantial fraction of them.  

Astronomers have begun to attempt to deconvolve the distribution of Rossiter-McLaughlin measurements into a mixture of systems formed through disk migration, planet scattering and secular perturbations \cite{MortonJohnson11b}.  However, we caution that there may be systematic biases in the outputs of such analyses, due to uncertainties in the treatment of tidal circularization.  Another potential confounding factor is the possibility of a primordial star-disk misalignment \cite{Batygin12,Lai12}.  Finally, we caution that an apparent trend of obliquity with stellar temperature \cite{Winn10} calls into question even qualitative predictions of tidal theory.  While the details remain unclear, the Rossiter-McLaughlin observations demonstrate that simple disk migration is inadequate to explain all hot-Jupiters.  Of course, these observations do not preclude disk migration from having operated in systems that were later sculpted by planet scattering or secular perturbations.  Indeed, planet scattering is more efficient at forming hot-Jupiters, if migration were to bring planets to ~1 AU prior to scattering \cite{MoorheadAdams05} than if scattering commenced at several AU.  

Another key observational result is the realization that hot-Jupiters are seldom accompanied by additional planets close to their host star.  This was foreshadowed by radial velocity planet searches \cite{Wright11} and dramatically confirmed by \Kepler observations of transiting hot-Jupiters \cite{Steffen12a}, as these provide precise constraints on both small planets with orbital periods of weeks to months (via photometry) and low-mass planets in or near mean-motion resonances (via transit timing variations).  This result is consistent with the broad predictions of hot-Jupiter formation via eccentricity excitation plus tidal circularization, but in stark contrast to the predictions of disk migration models \cite{Narayan05,CresswellNelson06}.  Thus, the isolation of hot-Jupiters suggests that there is a strong upper limit to the fraction of hot-Jupiters formed via disk migration.  
Recent radial velocity follow-up of systems with hot-Jupiters has found long-term radial velocity accelerations in roughly half of the surveyed hot-Jupiters \cite{Knutson13}.  The location of the second-closest planet in systems with hot-Jupiters bolsters the hypothesis that hot-Jupiters may frequently commence scattering while at an orbital distance $\sim$1 AU.  

\subsection{Summary of Hot-Jupiter Formation}
In summary, planet formation from a gaseous disk likely leads to forming many planets on low-eccentricity orbits.  Initially, close encounters lead to collisions and increasing planet masses.  Once the planets become massive enough to eject bodies from the gravitational potential well of the star, ejections become more common.  The recoil from scattering planets leads to eccentricity growth of giant planets, especially in the outer regions of the planetary system.  The AMD of planets that effectively eject smaller bodies is redistributed among all the remaining planets of a planetary system.  This leads to further close encounters and collision or ejections, depending on the masses and distances of the planets involved.  This process repeats, gradually thinning the planetary system, so that the remaining planets have masses and spacings that result in an instability timescale comparable to the age of the planetary system.  In a small fraction of systems, either the chaotic interactions of a multi-body system and/or the secular interactions of highly inclined system lead to the innermost giant planet passing close enough to the host star that tidal interactions circularize its orbit, leading to the formations of a hot-Jupiter.  While the future hot-Jupiter is circularizing, it cleans out the inner solar system by scattering any rocky planets in the inner planetary system into the star or the outer regions of the planetary system \cite{Mandell07}.

\section{Giant Planets near Snow Lines}

As RV surveys gained sensitivity to giant planets with greater orbital periods, a second population of giant planets was identified at  orbital periods from $\sim$300 days to $\sim$4 years \cite{Wright11}.  While the number of RV-discovered planets per decade in orbital period decreases for greater orbital periods, the true rate for this population could remain constant or even increase since RV surveys are incomplete for greater orbital periods.  Giant planets with orbital periods $\sim$year may well have formed further out in the disk and migrated to their current location.  The distribution for orbital separation of this population appears to peak near the predicted location of the snow line.  One possible interpretation is that many giant planets migrate to near the water snow line, i.e., the location in the disk where the solid surface density increases due to condensation of H$_2$O ice.  Since the snow line affects the formation of planetesimals, migration of giant planets towards the snow line could be accommodated by a variety of migration models, including migration through a gaseous disk, migration via planetesimal scattering or even via scattering of multiple planets or planetary cores.  

At orbital periods intermediate between that of the hot-Jupiters and giant planets near the snow line, RV surveys have revealed a paucity of giant planets with orbital periods from $\sim$10-200 days.  One possible interpretation is that giant planets typically migrate through this range of periods quickly.  For migration to strand these planets, the disk would need to dissipate during the brief window of time when the planet is passing through these intermediate orbital distances.  Possible mechanisms include photoevaporation of a gaseous disk \cite{AlexanderPascucci12} or the migration of a massive (and potentially distant) planet through a mean motion resonance exciting its eccentricity and destabilizing the planetesimal disk (e.g., Nice model)\cite{Tsiganis05}.  Either model requires some degree of tuning of parameters to strand a planet at intermediate orbital periods.  In principle, this could help explain the low abundance of giant planets at intermediate periods relative to either shorter or longer orbital periods.  

\section{Long-Period Giant Planets}

\subsection{Radial Velocity}
RV surveys have begun to discover giant planets with orbital distances ranging from a few to several AU \cite{Wright11}.  Many of these giant planets have significant eccentricities, suggesting that they may have previously ejected planets from their host planetary system.  Several studies have shown that planet-planet scattering naturally produces a broad distribution of eccentricities consistent with that observed \cite{Ford01,MarzariWeidenschilling02,MoorheadAdams05,Chatterjee08,FordRasio08,JuricTremaine08,Raymond10,Matsumura10}.  While this finding is robust to the choice of initial conditions, it also implies that it may not be possible to invert the observed eccentricity distribution to infer the architectures of young planetary systems shortly after the dissipation of the proto-planetary disk.  Thus, testing models of planet formation will require more detailed observations and comparisons.  
For example, the mass-period-eccentricity distribution may provide additional clues \cite{Raymond10}, if properly interpreted to account for observational selection effects.  Based on radial velocity observations, there appears to be a trend for more massive planets to have lower eccentricities than less massive planets \cite{FordRasio08}.  This is opposite the prediction of the simplest planet-planet scattering models.  However, the apparent trend could be easily accommodated if planet scattering often occurs while there is still a significant planetesimal disk which could damp eccentricities following the final scattering.  In this scenario, systems with more massive scattering planets would be less affected by the planetesimal disk and retain larger eccentricities than systems in which the planet-planet scattering involved less massive planets.  Thus, the mass-period-eccentricity correlation may provide a path towards constraining the late stage evolution of planetary systems.  However, one should keep in might that eccentricity measurements have a positive bias and the size of the bias increases for planets with radial velocity amplitudes small relative to the measurement precision \cite{ShenTurner08,Zakamska11}.  

Even more powerful constraints on planet formation are likely to come from the architecture and dynamical state of multiple planet systems (e.g., resonances, near resonances, amplitudes of secular modes).  
%
%
%
Unfortunately, characterizing the architectures of outer planetary systems is likely to prove significantly more difficult than planets with orbital periods of a few years or less.  Only several dozen stars have been observed at high precision long enough to characterize a Jupiter-clone.  Even for these targets, Saturn, Uranus and Neptune would appear only as small long-term trends given the duration of radial velocity planet searches.  Since an observed long-term acceleration might be due to a combination of multiple long-period planets, radial velocity surveys are more useful for placing limits on what planets with periods of decades or more are not present than for characterizing outer planetary systems.  

\subsection{Microlensing}
While direct imaging planet searches are just beginning to detect planets, microlensing surveys suggest that there is a substantial population of giant planets that has not been detected by radial velocity or transit searches \cite{Sumi11}.  These planets could either reside at large orbital separations from their host star or be free-floating after planet scattering ejected them from their original parent star.  In practice, both types of planets likely contribute to the microlensing population \cite{Veras09}.  Direct imaging searches may help disentangle this degeneracy.  

\subsection{Direct Imaging}
Recently, direct imaging planet searches have begun to detect low-mass companions with separations of tens to hundreds of AU (e.g., HR 8799)\cite{Marois08}.\footnote{It is often unclear whether these should be considered planets or brown dwarves, given our lack of knowledge of their mass or formation mechanism.  Here, we refer to all such companions as planets for simplicity, irrespective of their actual mass or formation history.}    
It is unclear whether these planets formed near their current separations or formed closer to their host star and migrated to their current location \cite{Dodson-Robinson09,Kratter10}.  Migration via a gas disk would require an unusually large disk mass at large separations.  This is an even more severe issue for planetesimal-driven migration.  Another possibility is that these planets may have migrated via planet-planet scattering, if these stars host at least one planet even more massive that the planet(s) observed at large separations \cite{ScharfMenou09,Veras09}.  In systems with extensive disks, the outer disk may have raised the planet's pericenter so that it decoupled from the inner regions of the planetary system.  Alternatively, for the young stars typically targeted by direct imaging searches, it is expected that we will detect planets that are still be in the process of being ejected from their star \cite{Veras09}.  

A final complication is that planets in wide orbits ($\ge$100 AU) could have formed around another star and exchanged into their present orbital configurations \cite{Malmberg11,Boley12}.  This does not offer an easy explanation for giant planets at very large separations, since typical exchange interactions are rapid and the ratio of the planet's final and initial semi-major axes only changes by a factor comparable to the ratio of host star masses.  Thus, exchanges may be important for explaining individual planetary systems (e.g., PSR 1620-26)\cite{Ford00a}, but are unlikely to explain the overall frequency of planets on very wide orbits.  

\section{Super-Earths \& Mini-Neptunes}

Over the past several years, radial velocity surveys began to discover planetary systems with Neptune and super-Earth-mass planets at short orbital periods \cite{Howard10}.  NASA's Kepler mission has demonstrated that such planets are much more common than giant planets and that most solar-type stars harbour a sub-Neptune-size planet \cite{Batalha13}.  Of more than 3,600 planet candidates identified by Kepler, only a small fraction have been dynamically confirmed (i.e., by radial velocity observations or transit timing variations).  However, detailed analyses of possible astrophysical false positives demonstrate that the average rate of false positives is sufficiently low that one can already begin to study the population of planet candidates as a whole \cite{MortonJohnson11a,Fressin13}, as long as one keeps in mind that  some subsets of planet candidates (e.g., giant planets with orbital periods less than $\sim~2$ days or $\sim~10-100$ days) have a higher rate of false positives \cite{Colon12,Santerne12}.  Fortunately, other important subsets, such as Neptune-size planets, can be shown to have a smaller false positive rate \cite{MortonJohnson11a,Fressin13}.  One very important subset of highly reliable planet candidates is those in systems with multiple transiting planet candidates \cite{Lissauer12}.
These can be considered planets with a confidence similar to that of planets discovered by radial velocity searches.  This allows one to conclude that planets with sizes between that of Earth and Neptune and orbital periods of weeks to months are the typical outcome of planet formation around solar-type stars.

\section{Short-Period Tightly-packed Inner Planetary Systems}

Beyond the sheer abundance of sub-Neptune-size planets, one particularly striking discovery is the abundance of systems with multiple transiting planets \cite{Lissauer11b,Fabrycky14}.  The typical system contains transiting planets with short orbital periods, ranging from $\sim$1-100 days.  The decrease in frequency at smaller orbital periods is real, while the decrease at large orbital periods may be at least partially due to detection biases, since both the geometric transit probability is lower and the signal to noise is smaller due to the smaller number of transits observed by Kepler.  
Such systems are typically tightly-packed, meaning that the orbital periods of planets transiting a common star are correlated with one another and not consistent with being drawn independently from a common orbital period distribution function.  While orbital periods range from hours to hundreds of days, the distribution of orbital period ratios is broadly peaked from $\sim$1.3 to 3.  Therefore, we refer to these systems as Short-Period Tightly-packed Inner Planetary Systems (STIPS).  

While theorists have developed multiple mechanisms for migrating giant planets from beyond the snow line to near their present location, the lower masses of sub-Neptune-size planets mean that {\em in situ} formation can not be dismissed so easily \cite{ChiangLaughlin13}.  The mass of most such planets is likely dominated by rock, ices or water, but not gas, even for low-density, Neptune-size planets.  Therefore, these planets need not accrete a substantial rocky core before the protoplanetay disk is cleared.  Indeed, it is conceivable that many of these atmospheres are primarily the result of outgassing, rather than accretion of gas from the disk.  The combination of lower masses and the possibility of a longer timescale for formation makes {\em in situ} formation worthy of serious consideration.  

Of course, just because theory does not prohibit in situ formation, one should not assume that planets in STIPS did not undergo significant migration.  Since sub-Neptune-size planets would not clear a gap, they are expected to migrate through a given gaseous disk more rapidly than giant planets.  While early work suggested that planetary migration would accelerate as planets spiral closer to their host stars, recent studies incorporating more realistic disk models predict material may accumulate or become trapped at orbital distances that are not dissimilar from the location of Kepler's sub-Neptune-size planets \cite{PaardekooperPapaloizou09}.  
Multiple mechanisms have been proposed for for setting the overall scale for the orbital distances of STIPS, including the transition to an MRI-active inner region \cite{ChatterjeeTan13} and the silicon sublimation front \cite{ChiangLaughlin13,BoleyFord13}.  Once one planet has formed at small distances, it can perturb the disk structure so that additional planets form or become trapped in resonances, not far behind \cite{Kobayashi12}.  Subsequent dynamical evolution can lead to spreading out of planetary systems \cite{HansenMurray12} and disruption of resonances.  

\subsection{Orbital Spacing of Planets in STIPS}

While dynamical stability prohibits systems of two giant planets with a period ratio less than $\sim$1.3 (excluding small islands of stability near mean motion resonances), Kepler's systems of small and presumably lower-mass planets can remain stable even for much tighter spacings such as $\sim$1.17 (Kepler-36)\cite{Carter12,Deck12}.  
The relative frequency of planetary systems with one, two and three transiting planets implies that there is a population of STIPS with low mutual orbital inclinations, since systems with high mutual inclinations would rarely result in multiple transiting planets \cite{Lissauer11b}.  Adding information about the ratios of transit durations shows that these systems also have low eccentricities \cite{FangMargot13,Fabrycky14}.   
The sizes of adjacent transiting planets in STIPS are observed to be correlated with each other \cite{Ciardi13}.  
These observations are qualitatively consistent with both models of {\em in situ} formation and formation at larger separations followed by migration to their currently observed orbits.  

While Kepler detects STIPS with two transiting planets more often than higher multiplicity systems, most of this trend is a detection effect due to the geometric detection probability.  Therefore, the typical outcome of planet formation appears to include forming multiple, tightly-packed, sub-Neptune-size planets with similar sizes and orbital separations.  Most such planetary systems are not in mean motion resonances (MMRs)\cite{VerasFord12}, but a significant fraction of such systems have ratios of orbital periods just slightly larger than that of a first-order MMR (typically 2:1, 3:2, or 4:3)\cite{Fabrycky14}.  This presents a significant challenge to planet formation theories.  Clearly, MMRs are playing a role in the planet formation process.  However, a slow and smooth migration would be expected to trap planets in mean motion resonances.  This has led some theorists to propose that pairs of planets are first trapped into MMRs and then dissipation causes the planets to be removed from resonance.  However, the amount of dissipation required to match the observed period ratios is so large that it implies uncomfortably large tidal quality factors \cite{LithwickWu12} or rapid eccentricity dissipation timescales \cite{GoldreichSchlichting13}.  An alternative model which invokes mass growth rather than migration or dissipation appears to require masses much larger than is plausible given the typical planet sizes \cite{Petrovich13}.  Preliminary work suggests that  interactions with a planetesimal disk may be important for removing planets from MMRs and leaving them with the orbital period ratios similar to those observed by Kepler. 

\subsection{Transit Timing Variations in STIPS}

In the case of a single star and a single planet, the orbit is strictly Keplerian and transit times follow a linear ephemeris.  In planetary systems with additional planets or stellar companions, the mutual gravitational interactions cause the transit (or eclipse) times to deviate from a strict periodicity.  The differences between the actual times of transit and the times predicted by the best-fit linear ephemeris are known as transit timing variations (TTVs).  
In many of the STIPS with planets near a MMR, Kepler has measured significant TTVs \cite{Ford12b,Mazeh13}.  
Even in cases with low TTV signal-to-noise or a TTV timescale significantly greater than the timescale of observations, TTVs can be used to confirm the planetary nature of the planet candidates \cite{Holman10,Fabrycky12a,Ford12a,Steffen12b,Steffen12c,Steffen13,Xie13} and to provide constraints on planet masses and eccentricities \cite{Lithwick12,WuLithwick13,HaddenLithwick13}.  While TTVs (presumably due to non-transiting planets) are often detected for planets transiting stars with no other known transiting planet candidates \cite{Ford12a,Mazeh13}, only in rare cases do these yield unique orbital solutions \cite{Dawson12,Nesvorny12,Nesvorny13}.  Nevertheless, simply observing that the frequency of detectable TTVs increases significantly for systems with multiple transiting planet candidates \cite{Ford12a,Xie13} provides information about the frequency of non-transiting planets in such systems and thus the frequency of closely-spaced planets and the distribution of mutual orbital inclinations.  

The realization that STIPS systems are typical raises new questions about the nature and formation of this new class of planets  with sizes between that of Earth and Neptune and the architecture of STIPS. 
The most basic questions have to do with the masses, densities and compositions of such planets. 
Since many STIPS exhibit significant TTVs, these systems provide a critical information for characterizing the masses and eccentricities of small planets \cite{Lissauer11a,Carter12}.  Over the next several years, detailed TTV analyses of several specific systems will provide precise masses for small planets.  Additionally, a systematic study of systems with TTVs is likely our best tool for characterizing the masses and densities of a large sample of small planets and thus the planet mass-radius relationship more generally.

\begin{acknowledgments}
E.B.F. acknowledges many valuable discussions with Aaron Boley, Sourav Chatterjee, Dan Fabrycky, Robert Morehead and Darin Ragozzine, as well as the entire Kepler science team and particularly the Kepler multi-body and transit timing variations working groups.  
This material is based on work supported by the National Aeronautics and Space Administration under grants NNX08AR04G and NNX12AF73G issued through the Kepler Participating Scientist Program and grant NNX09AB35G issued through the NASA Origins of Solar Systems program.  
E.B.F. acknowledges the support of the University of Florida, The Pennsylvania State University, the Penn State Center for Exoplanets and Habitable Worlds and the Institute for CyberScience.  
\end{acknowledgments}

\end{article}

\end{document}